\documentclass[%
 reprint,
 amsmath,amssymb,
aip,
]{revtex4-1}
\setcitestyle{numbers,square}

\usepackage{physics}
\usepackage{braket}
\usepackage{graphicx}
\usepackage{dcolumn}
\usepackage{bm}
\usepackage[normalem]{ulem} 
\usepackage{color}

\usepackage{hyperref}
\hypersetup{
    colorlinks,%
    citecolor=blue,%
    linkcolor=blue,%
    urlcolor=blue
}

\newcommand{\Ni}{(NiC$_4$S$_4)_3$}
\newcommand{\fp}{\it first-principles}
\newcommand{\avg}[1]{\langle #1 \rangle}

\begin{document}

\title{Layertronic control of topological states in multilayer metal-organic frameworks}

\author{F. Crasto de Lima} 
\email{felipe.lima@ufu.br}
\author{G. J. Ferreira}
\author{R. H. Miwa}

\affiliation{Instituto de F\'isica, Universidade Federal de Uberl\^andia, \\ C.P. 593, 38400-902, Uberl\^andia, MG,  Brazil}

\date{\today}

\begin{abstract}

{We investigate the layer localization control of two-dimensional states in multilayer metal-organic frameworks (MOFs). For finite stackings of {\Ni} MOFs, the weak van der Waals coupling between adjacent layers leads to a Fermi level dependent distribution of the electronic states in the monolayers. Such distribution is reflected in the topological edge states of multilayer nanoribbons. Moreover, by applying an external electric field, parallel to the stacking direction, the spacial localization of the electronic states can be controlled for a chosen Fermi energy. This localization behavior is studied comparing density functional theory calculations with a kagome lattice tight-binding model. Furthermore, for infinite stacked nanoribbons, a new V-gutter Dirac state is found in the side surfaces, which allows anisotropic current control by tuning the Fermi energy. Our results can be immediately extended to other kagome MOFs with eclipsed stackings, introducing a new degree of freedom  (layer localization) to materials design.}

\end{abstract}

\maketitle

\section{Introduction}

Two-dimensional topological insulators (2DTI) have attracted broad interest for novel applications due to their characteristic dissipationless transport properties \cite{RMPSinova2015, JPCLKou2017}. While graphene was the first predicted 2DTI, its topological gap is too small to be observed experimentally \cite{PRLKane2005,PRBYao2007}. Recently, a new class of large gap 2D topological materials, the metal-organic frameworks (MOFs), emerged as an ideal playground to achieve different properties. That is mainly due to a rich tunability of MOFs via different combinations of metals and organic ligands to form designed materials with selected properties \cite{ChemCommunMiguel2016, NATCHEMColson2013, PCCPChen2015, LANGMaeda2016, CCRZhao2018,  PCCPGutzler2016}. For instance, it has been shown that Mn, Fe and Co-bis(dithiolene) are ferromagnetic half-metals, while n-doped Ru-bis(dithiolene) has a large gap Quantum Anomalous Hall phase \cite{JPCCChakravarty2016,PCCPdeLima2018}. On the other hand, by keeping iron as the metal and changing the organic ligand to a coronene based molecule, a ferromagnetic semicondutor phase has been found \cite{NatureDong2018}. Currently, a great number of MOFs are predicted to host topological phases \cite{PRLLiu2013, PRB90Zhao, NanoscaleZhou2015, NanoLettersZhang2016, PRBYamada2016, PRBZhang2016, APLHsu2018, NanoscaleNi2018, JPCCSun2018}. Within these materials, nickel and palladium-bis(dithiolene) monolayers were experimentally synthesized \cite{JACSSakamoto2013, CPCSakamoto2015}, and theoretically shown to host a 2DTI phase for n-doped systems \cite{NANOFeng2013}. Although the monolayer system need to be n-doped to achieve the 2DTI phase, in a recent work we have shown that for bilayers of nickel and platinum-bis(dithiolene), the topological gap falls onto the Fermi level \cite{PRBdeLima2017}.
	
The development of novel nano-devices based on 2DTIs is linked to the experimental control of its properties, and manipulation of its charge, spin and valley degrees of freedom \cite{SCIENCEWolf2001, NATURESchaibley2016}. Recently, it has been proposed a bending strain perturbation approach to control spin currents in 2DTIs \cite{NatCommHuang2017}. Meanwhile, ensuing research has shown layer localization control of topological states by an applied external electric field in MOF bilayer nanoribbons \cite{PRBdeLima2017} and bilayer graphene grain boundaries \cite{2DMJaskolski2018}. Furthermore, the exploration of van de Waals heterostructures allows for designed properties by combining the stacking of two-dimensional materials \cite{NatureGeim2013,SCIENCENovoselov2016}. Within the heterostructures, the number of stacked layers and the stacking order has great impact in the electronic properties \cite{PRBKim2011, SCIREPAraujo2018, 2DMSuchun2018}. Therefore, in order to control the electronic properties of MOFs multilayer, a clear understanding of its stacking is required.

In this paper, we study the stability and control of the electronic properties of the MOF {\Ni} stacking, from a monolayer to the bulk limit. Our {\fp} results show that the mirror symmetric AA stacking is the most stable configuration. For a finite AA stacking, each layer corresponds to a 2DTI system with helical edge states weakly coupled to its counterpart in adjacent layers. Here, the edge states are obtained from a tight-binding model of the stacking, fitted to match the {\fp} data. These features allow us to propose a novel electric field control of the layer localization of the edge states. For an infinite AA stacking, in the bulk limit, the edge states from each layer combine to form a topological V-gutter-like surface states dispersion along the stacking direction. These surface states are highly anisotropic and tunable, introducing a new feature to the {design of future multifunctional spintronic devices \cite{ADVMATGuo, NATURESoumyanarayanan2016}.}

\section{Results and Discussions}

\subsection{Stacking geometry}
  
\begin{figure}
\includegraphics[width=\columnwidth]{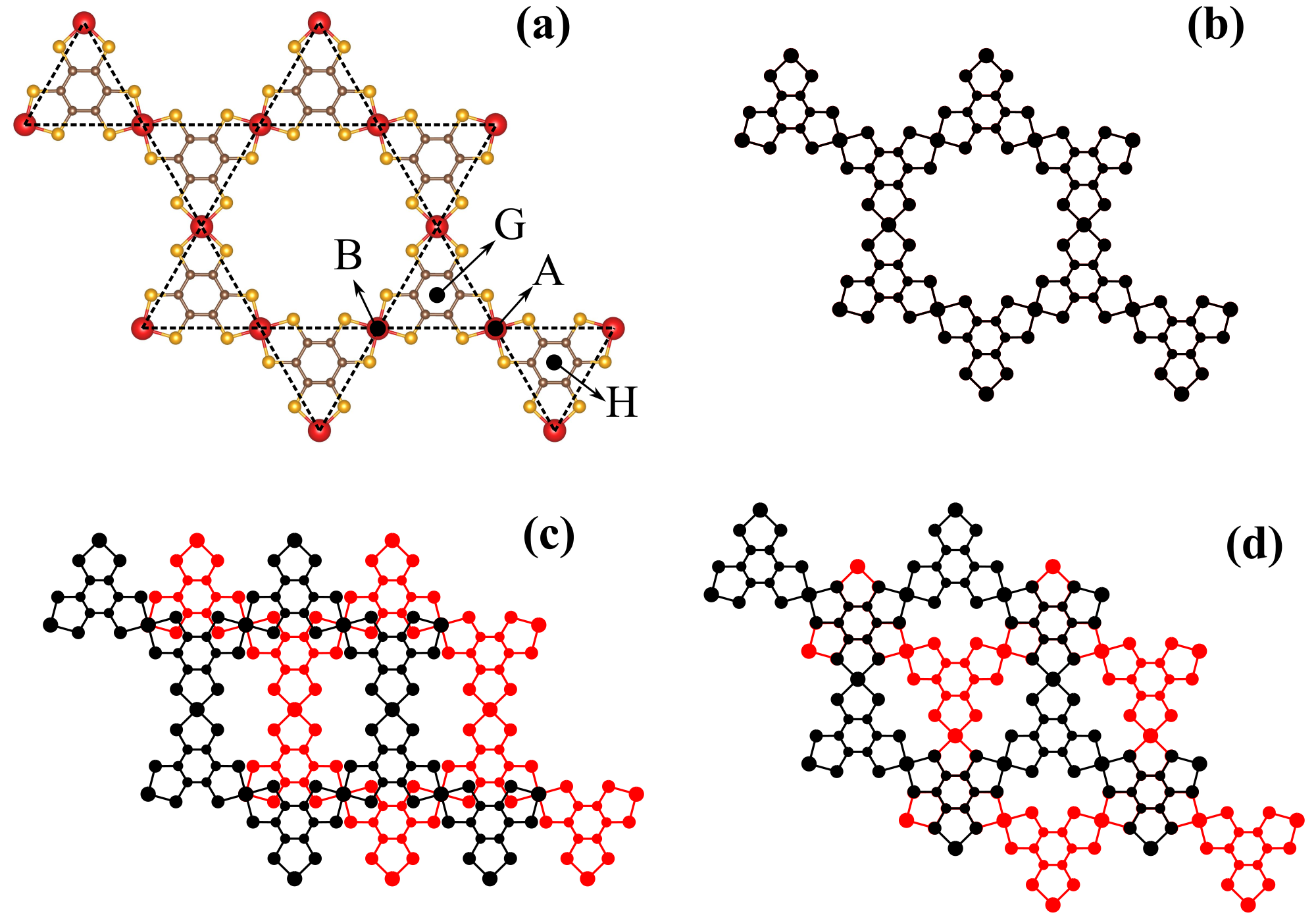}
\caption{\label{fig:str} MOF {\Ni} monolayer atomic structure (a) with C, S and Ni atoms in brown, orange and red, respectively. Staking geometry for the eclipsed AA (b) AB (c) and GH (d), with the top layer in black, and bottom layer in red.}
\end{figure}

The Nickel bis(dithiolene) monolayer presents a planar geometry, composed by Ni atoms connected by carbon-sulfur molecules (C$_6$S$_6$), which combine to form a kagome lattice [dashed lines in Fig.\,\ref{fig:str}(a)]. In order to analyze the energetic stability of multilayered {\Ni} systems, we have considered various stacking geometries. For instance, in {\Ni} bilayers (2ML), the stacking geometry is characterized by the alignment of sites X and Y of the different layers (2ML-XY),  with X and Y = A, B, G, and H, as indicated in Fig.\,\ref{fig:str}(a). Within this nomenclature, 2ML-AA [Fig.\,\ref{fig:str}(b)] represents the eclipsed structure, 2ML-AB [Fig.\,\ref{fig:str}(c)] dislocates one layer along the zigzag direction, and 2ML-GH [Fig.\,\ref{fig:str}(d)] mimics the Bernal stacking of graphene.

In a previous study \cite{PRBdeLima2017}, we have found the 2ML-AA configuration to be the most stable. There, the calculations were performed using the non-local  vdW-DF2 \cite{PhysRevB.82.081101} correction to describe the vdW interactions. Indeed, the energetic preference of 2ML-AA was also verified in Refs.~\cite{JPCCShojaei2014, JPCMChacham2017} using different approaches to describe the long range vdW interactions \cite{grimmeJCompChem2006, PhysRevLett.92.246401, klimesJPhysC2010}. In contrast, while Ref.~\cite{2DMSuchun2018} shows an energetic preference for the 2ML-GH configuration using the vdW-D3 \cite{grimme2010consistent}, our results with this functional corroborates the previous works, showing a preference for the 2ML-AA case.

Notice that, despite the weak chemical interactions (essentially vdW) between the stacked layers, the electronic band structure of multilayered {\Ni} systems strongly depends on the stacking geometry. Indeed, the 2ML-AA and -GH configurations present significantly different band structures near the Fermi level, see Appendix \ref{app:band-stacking}. Moreover, here we are interested in the topological insulator properties of {\Ni} MOFs \cite{NANOFeng2013}, and such a dependence  will reflect on the characteristics of the edge states in bilayer \cite{PRBdeLima2017}, and multilayer {\Ni}.
  
A detailed analysis of the binding energy $E^b$ \footnote{The binding energy ($E^b$) of {\Ni} bilayer systems is defined as, $ E^b = E^{\rm ML} - \frac{1}{2}E^{\text{\rm 2ML-XY}}, $ where $E^{\text{\rm 2ML-XY}}$ and $E^{\rm ML}$  are the total energies of the {\Ni} 2ML-XY, and (free standing) monolayer. Similar definition was applied  for the 3ML-XYZ and bulk-XY systems.} and geometric properties of different stackings is shown in Table\,\ref{tab:par}. Here, the total energy calculations were performed considering full relaxation of the atomic positions and lattice vectors, until the force in each atom was less than 5 meV/{\AA}. Moreover, we compare results obtained using different functionals, {\it viz.}: LDA \cite{Pe}, GGA-PBE \cite{PBE}, local vdW-D3 correction \cite{grimme2010consistent}, and  the non-local vdW-DF2 correction \cite{PhysRevB.82.081101}. The wave functions were expanded in a plane-wave basis with cutoff energy of 450 eV. For BZ integration a $4\times4\times6$ ($4\times4\times1$) k-mesh for bulk (finite stacking) systems was considered.
{We have also considered the GGA+U method to correct the on-site repulsion of Ni d-orbitals. As shown in Appendix \ref{app:DFT+U}, these corrections do not affect significantly our results.}


\begin{table}[h]
\caption{\label{tab:par}Lattice parameter $a$ ({\AA}), averaged inter-layer 
distance $d$ ({\AA}), corrugation ${\delta z} = \sqrt{\avg{z^2} - 
\avg{z}^2}$ ({\AA}) and binding energy ($E^b$ in eV/layer) of the bulk 
and layered systems.}
\begin{ruledtabular}
\begin{tabular}{c|cccc}
\multicolumn{5}{c}{PBE} \\
\hline
System &$a$&$d$&$\delta z$&$E^b$\\ 
\hline
1ML     & 14.62 &  -   & 0.00 & 0.000 \\
2ML-GH  & 14.63 & 4.23 & 0.01 & 0.037 \\ 
2ML-AB  & 14.62 & 3.99 & 0.05 & 0.033 \\
2ML-AA  & 14.62 & 4.69 & 0.01 & 0.022 \\ 
3ML-AAA & 14.64 & 3.95 & 0.00 & 0.004 \\ 
bulk-GH & 14.63 & 3.79 & 0.00 & 0.055 \\ 
bulk-AB & 14.63 & 3.73 & 0.00 & 0.002 \\ 
bulk-AA & 14.63 & 3.93 & 0.00 & 0.009 \\ 
\hline
\multicolumn{5}{c}{LDA} \\
\hline
System &$a$&$d$&$\delta z$&$E^b$\\ 
\hline
1ML     & 14.37 &  -   & 0.00 & 0.000 \\ 
2ML-GH  & 14.42 & 2.96 & 0.11 & 0.468 \\ 
2ML-AB  & 14.36 & 3.16 & 0.06 & 0.429 \\ 
2ML-AA  & 14.44 & 2.96 & 0.08 & 1.051 \\ 
3ML-AAA & 14.43 & 3.11 & 0.04 & 1.138 \\ 
bulk-GH & 14.39 & 3.13 & 0.00 & 0.985 \\ 
bulk-AB & 14.40 & 3.29 & 0.00 & 0.888 \\ 
bulk-AA & 14.45 & 3.12 & 0.00 & 1.800 \\ 
\hline
\multicolumn{5}{c}{vdW-D3} \\
\hline
System &$a$&$d$&$\delta z$&$E^b$\\ 
\hline
1ML     & 14.62 &   -  & 0.00 & 0.000 \\ 
2ML-GH  & 14.61 & 3.26 & 0.11 & 0.679 \\ 
2ML-AB  & 14.60 & 3.34 & 0.15 & 0.661 \\ 
2ML-AA  & 14.64 & 3.54 & 0.01 & 0.860 \\ 
3ML-AAA & 14.64 & 3.57 & 0.00 & 1.206 \\ 
bulk-GH & 14.62 & 3.31 & 0.00 & 1.501 \\ 
bulk-AB & 14.60 & 3.46 & 0.00 & 1.404 \\ 
bulk-AA & 14.65 & 3.35 & 0.00 & 1.968 \\ 
\hline
\multicolumn{5}{c}{vdW-DF2} \\
\hline
System &$a$&$d$&$\delta z$&$E^b$\\
\hline
1ML     & 14.83 &   -  & 0.00 & 0.000 \\ 
2ML-GH  & 14.83 & 3.33 & 0.11 & 0.645 \\
2ML-AB  & 14.82 & 3.34 & 0.21 & 0.602 \\
2ML-AA  & 14.86 & 3.64 & 0.01 & 0.803 \\ 
3ML-AAA & 14.85 & 3.65 & 0.00 & 1.107 \\ 
bulk-GH & 14.84 & 3.46 & 0.00 & 1.360 \\ 
bulk-AB & 14.82 & 3.56 & 0.00 & 1.267 \\
bulk-AA & 14.87 & 3.56 & 0.00 & 1.744 \\
\end{tabular}
\end{ruledtabular}
\end{table}

Starting with the 2ML case, the PBE results show larger interlayer distances $d$, and lower values of binding energies when compared with the other functionals, thus indicating that the vdW interactions play an important role on the energetic stability and the equilibrium geometry of the multilayered {\Ni} systems. When vdW corrections are included, we find that 2ML-AA stackings are more stable than 2ML-GH \footnote{Although our calculations have {more base functions and denser k-grid} than that of Ref.~\cite{2DMSuchun2018}, we have also run calculations with exact the same parameters listed in this reference, but even in this case we find that the 2ML-AA stacking is more stable.}, with a $E^b$ difference of $\sim$0.15\,eV/layer. The 2ML-AA stacking shows $E^b =$ 0.80 (0.86)\,eV/layer, and interlayer distance $d=$ 3.64 (3.54)\,{\AA}, within the vdW-DF2 (vdW-D3) functional. In general, the bond distances obtained from the vdW-DF2 approach are larger than those obtained by the vdW-D3 approximation. Nonetheless, both vdW corrections capture the enhancement of the vertical corrugation ($\delta z$) due to the reduction of the interlayer distance in 2ML-AB, $\delta z$=0.1$\rightarrow$0.21  and 0.1$\rightarrow$0.15\,{\AA}, for vdW-DF2 and -D3, respectively. The energetic preference for the 2ML-AA configurations has been also verified within the LDA approximation. However, a comparison between the LDA and vdW results show that, within the LDA approach, the $E^b$ difference between 2ML-AA and -AB stackings is larger (0.62\,eV/layer), and the 2ML-AA configuration shows a smaller interlayer distance (2.96\,{\AA}), which leads to a larger corrugation $\delta z$=0.08\,{\AA}.

For trilayers and the bulk systems, further binding energy results, shown in Table \ref{tab:par}, also indicate the energetic preference for the AA stacking (3ML-AAA and bulk-AA). In comparison with the 2ML-AA case, the binding energy of the bulk-AA system increases by $\sim$1\,eV/layer, followed by a reduction of the interlayer distance by  0.08\,{\AA}, $d$=3.64$\rightarrow$3.56\,{\AA},  (0.19\,{\AA}, 3.54$\rightarrow$3.35\,{\AA}) using the vdW-DF2 (vdW-D3) approximation. Within the LDA approach, we have also found the bulk-AA stacking to be the most stable. However, in contrast with the (vdW) results above, at the equilibrium geometry the interlayer distance increases by 0.16\,{\AA} when compared with the bilayer 2ML-AA, 2.96$\rightarrow$3.12\,{\AA}. These findings suggest that the vdW and LDA approximations present the same qualitative picture regarding the energetic stability of multilayered {\Ni} systems. However, the LDA approach fails to describe their equilibrium geometries. On the other hand, the PBE functional fail to describe the interlayer interaction, with the interlayer distance overestimated and smaller binding energies. Hereafter we will consider the most stable eclipsed (AA) geometry, obtained using the vdW-DF2 approach.

\subsection{Electronic properties}

\begin{figure*}
\includegraphics[width=\textwidth]{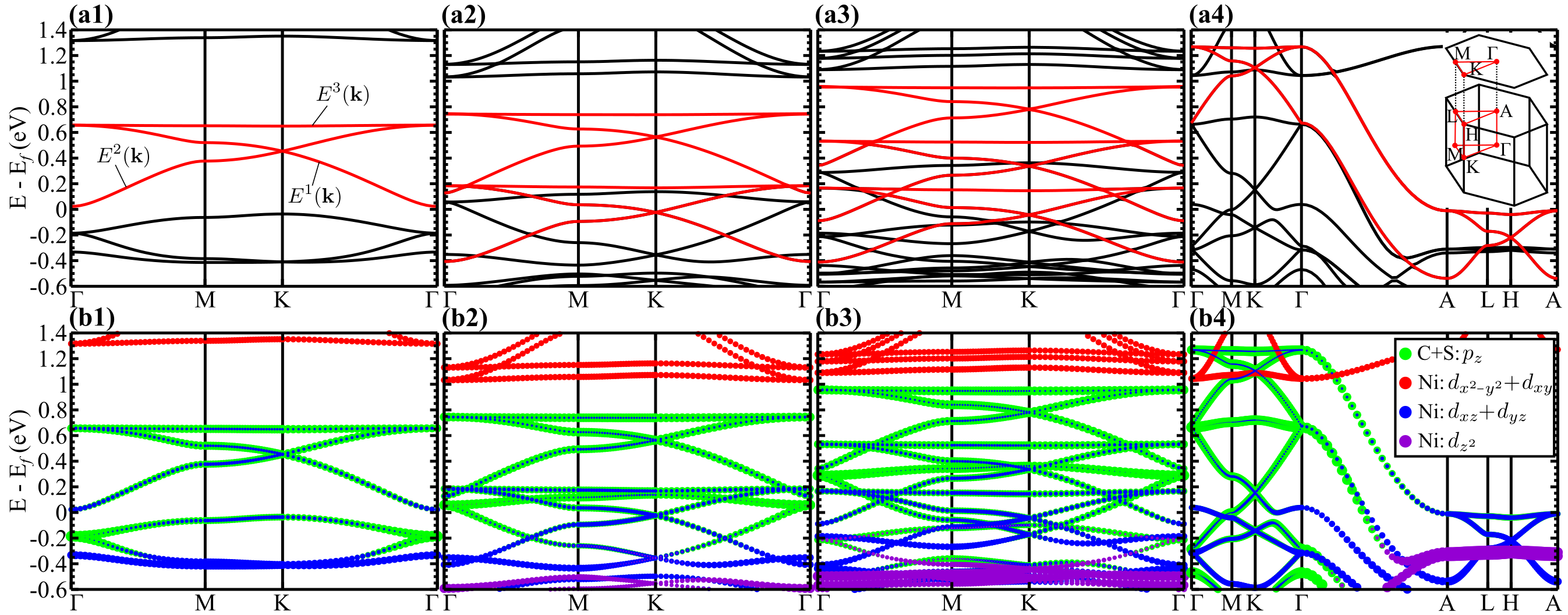}
\caption{\label{fig:band} DFT band structure without SOC for 1ML (a1) 2ML (a2) 3ML (a3) and bulk (a4) {\Ni} stacking. The red lines indicate the kagome bands. In (b1)-(b4) the projected orbital contributions are show on top of the band structures from the top panels.}
\end{figure*}

To analyze the electronic structure of the AA stacking, we first consider the DFT results without SOC, as shown in Fig.~\ref{fig:band} for 1ML, 2ML, 3ML, and bulk {\Ni} systems. Later on, in Section \ref{sec:edge}, we will introduce the SOC to discuss the topological edge states. 

For the single layer {\Ni}, the typical kagome bands (KBs), indicated by red lines in Fig.\,\ref{fig:band}(a1), are characterized by two energy bands [$E^1(\bm k)$ and $E^2(\bm k)$] with linear dispersion at the K-point, yielding a Dirac cone, connected to a nearly flat band [$E^3(\bm k)$] at the $\Gamma$-point. The orbital projection, Fig.\,\ref{fig:band}(b1), shows that the KBs are {mostly composed by C-p$_z$ and S-p$_z$ orbitals, while near the center of the Brilloin zone, and along the M-K direction, the Ni-d$_{xz}$ and -d$_{yz}$ orbitals also contribute to the formation of those energy bands.} For the 2ML-AA case, one finds two sets of KBs, Fig.\,\ref{fig:band}(a2). There, one set of KBs is fully unoccupied, while the other is partially occupied, with the Dirac point lying slightly below the Fermi level. Such a metallic character is also seen in 3ML, Fig.\,\ref{fig:band}(a3), where three sets of KBs are seen. The orbital projection on these energy bands, Figs.\,\ref{fig:band}(b2) and \ref{fig:band}(b3), reveal that, despite showing similar energy dispersion, the KBs present distinct orbital distribution. Namely, the contribution of the Ni-d$_{xz}$/d$_{yz}$ [blue circles in Fig.\,\ref{fig:band}(b1)-(b3)] orbitals are larger on high energy KBs, whereas the C-p$_z$ and S-p$_z$ orbitals [green circles in Fig.\,\ref{fig:band}(b1)-(b3)] are dominant on the low energy KBs. Such an orbital distribution has been strengthened in the {\Ni} bulk system. As shown in Figs.\,\ref{fig:band}(a4) and \ref{fig:band}(b4), the contribution of the Ni-d (C-p$_z$ and S-p$_z$) orbitals becomes dominant on the fully occupied (empty) KBs at high  (low) energy, which are localized along the A-L-H-A ($\Gamma$-M-K-$\Gamma$) directions, see inset of Fig.\,\ref{fig:band}(a4).

\subsubsection{Effective tight-binding model}
{Within the eclipsed AA stacking geometry, the interlayer coupling is mostly given by weak vdW interactions between kagome sites aligned along the stacking direction $z$. Given this scenario, we propose a tight-binding (TB) model, for the stacking of $N$ layers, considering first neighbors coupling between monolayers. The resulting Hamiltonian reads}
\begin{equation}
H_N = h_{3}({\bm k}) \otimes \mathbb{I}_{N} + \Delta \; \mathbb{I}_{3} \otimes \sum_{j=1}^{N-1} \ket{j+1} \bra{j} + c.c.,
\end{equation}
where $\ket{j}$ labels the subspace of each layer $j$, $\mathbb{I}_{N}$ is a $N\times N$ identity matrix within this subspace{, and ${\bm k} = (k_x,\,k_y)$ is the in-plane momentum. The $N$-layer Hamiltonian $H_N$ above is written in terms of the monolayer's $3\times3$ Hamiltonian $h_{3}({\bm k})$, which gives the characteristic KBs dispersion. The vdW interlayer coupling is set by $\Delta$, for which the diagonal form ($\mathbb{I}_{3}$) labels the direct coupling between equivalent atoms in the AA stacking geometry.} The second term of $H_N$ is a tridiagonal Toeplitz matrix \cite{trimatrix}, yielding a closed form for the eigenvalues as
\begin{equation}
\lambda^{n}_{\nu}({\bm k}) = E^n({\bm k}) - 2\,|\Delta|\, \cos \left( 
\frac{\nu \pi}{N+1} \right), \label{eigenval}
\end{equation}
with $E^n(\bm k)$ being the kagome monolayer dispersion [Fig.\ref{fig:band}(a1)], where $n$ labels each of the three KBs for each set $\nu = 1, \dots, N$. The eigenvectors are $\ket{n,\,{\bm k}} \otimes \ket{\nu}$, where the first term is an eigenstate of $h_3(\bm{k})$. To understand the second quantum number $\nu$ and the eigenstate component $\ket{\nu}$, it is interesting to notice that the tridiagonal Toeplitz form of the $\Delta$ coupling is equivalent to the Hamiltonian of a scalar $N$-sites chain, \textit{i.e.,} a quantum well along the stacking direction. Namely, $\ket{\nu} = \Omega\, [\phi_{1,\nu}, \cdots, \phi_{j,\nu}, \cdots, \phi_{N,\nu}]$, where $\Omega = \sqrt{2/(N+1)}$ is a normalization factor, and
\begin{equation}
\phi_{j,\nu} = \sin \left( \frac{j \nu \pi}{N+1} \right) \equiv \sin\left(k_\nu z_j\right)
\end{equation}
is the contribution of the state $\nu$ to the $j$-th layer. In the second form of $\phi_{j,\nu}$ above, we use the quantum well analogy to introduce $k_\nu = \nu \pi/L_N$ as the quantized $k_z$ momentum along the stacking direction $z_j = j d$, where $d$ is the interlayer distance, and $L_N = (N+1)d$ is the effective stacking length \footnote{In this analogy, the quantum well chain has $N+2$ sites distant by $d$, yielding a length $L_N = (N+1)d$. A hard-wall boundary condition makes $\phi_{0,\nu} = \phi_{N+1,\nu} = 0$, such that the real layers lie within $1 \leq j \leq N$.}. This analogy is useful to interpret the layer localization and the electric field effects below.

To obtain an estimate for $\Delta$, we compare the {\fp} band structures with the eigenvalues $\lambda^{n}_{\nu}({\bm k})$ of our TB model. The {\fp} energy dispersion for 2ML and 3ML, Fig.~\ref{fig:band}, show that each set of KBs are nearly rigidly shifted from one another. This matches the results of the TB model, where $\lambda^n_{\nu+1}({\bm k}) - \lambda^n_\nu({\bm k})$ is independent of $\bm{k}$. For 2ML [Fig.\,\ref{fig:band}(a2)], the two sets ($\nu = 1$ and 2) of KBs are distant by $\lambda^n_2({\bm k}) - \lambda^n_1({\bm k}) = 2|\Delta| \approx 0.56$~eV, thus $|\Delta| \approx 0.28$~eV. For the 3ML case we also find $|\Delta|\approx 0.28$\,eV.

Taking the bulk limit of $N\rightarrow\infty$ layers, we can explore the quantum well analogy to express $\lambda^{n}_{\nu}({\bm k}) \rightarrow \lambda^{n}_{\nu}({\bm k}, k_z) = E^n({\bm k}) - 2\,|\Delta|\, \cos(k_z d)$, assuming that now $k_z$ is a continuous variable, since $L_N \rightarrow \infty$. The $\cos(k_z d)$ dispersion is clearly seen in Fig.~\ref{fig:band}(a4) connecting the topmost and bottommost KBs along the $\Gamma$-A direction ($\parallel k_z$). From this dispersion, we obtain $|\Delta|=0.31$\,eV. Such an increase of $\Delta$ (0.28\,$\rightarrow$\,0.31\,eV), although small, is a result of the reduction of the interlayer distance in the bulk system with respect to the few layer systems, 2ML or 3ML in Table\,\ref{tab:par}.

\begin{figure}
\includegraphics[width=\columnwidth]{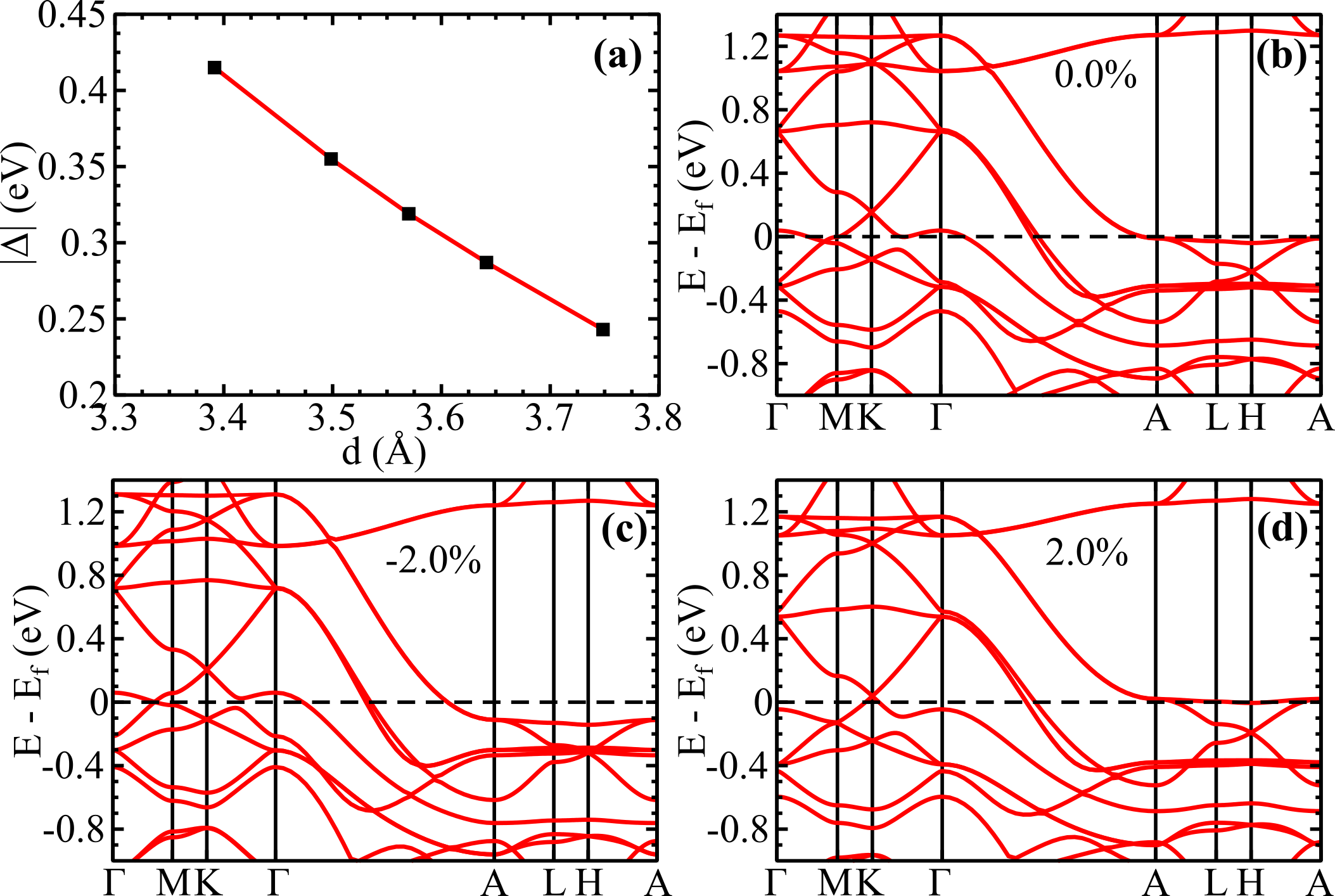}
\caption{\label{fig:bulk-strain} (a) Variation of coupling $|\Delta|$ with the  interlayer distance $d$. Illustrative band structure for the bulk system with equilibrium $d$ (b), with $d$ increased by $2\%$ (c), and decreased by $2\%$ (d).}
\end{figure}
 
To further explore the band width changes, and support our TB model, in Fig.\,\ref{fig:bulk-strain}(a) we show $|\Delta|$ as a function of the interlayer spacing $d$ for the {\Ni} bulk system. This distance can be controlled by an external pressure (uniaxial strain), thus allowing us to tune $\Delta$. Since the $\Delta$ coupling reflects the hybridization of neighboring layers, it decreases as $d$ increases. As discussed above, $\Delta$ is extracted from the bands connecting KBs along the $\Gamma$-A direction. At the equilibrium interlayer distance, $d$=3.56\,{\AA}, the energy difference between the upper- and bottommost KBs is about 1.24\,eV, Fig.\,\ref{fig:bulk-strain}(b). Meanwhile, for $d$=3.63 (3.49)\,{\AA}, this energy split reduces (increases) to 1.12 (1.44)\,eV [Figs.\,\ref{fig:bulk-strain}(c) and \ref{fig:bulk-strain}(d)]. That is, a $\sim 2$\% uniaxial strain induces a $\sim 10$\% change in $\Delta$.

\subsubsection{Layer localization}

\begin{figure}
\includegraphics[width=\columnwidth]{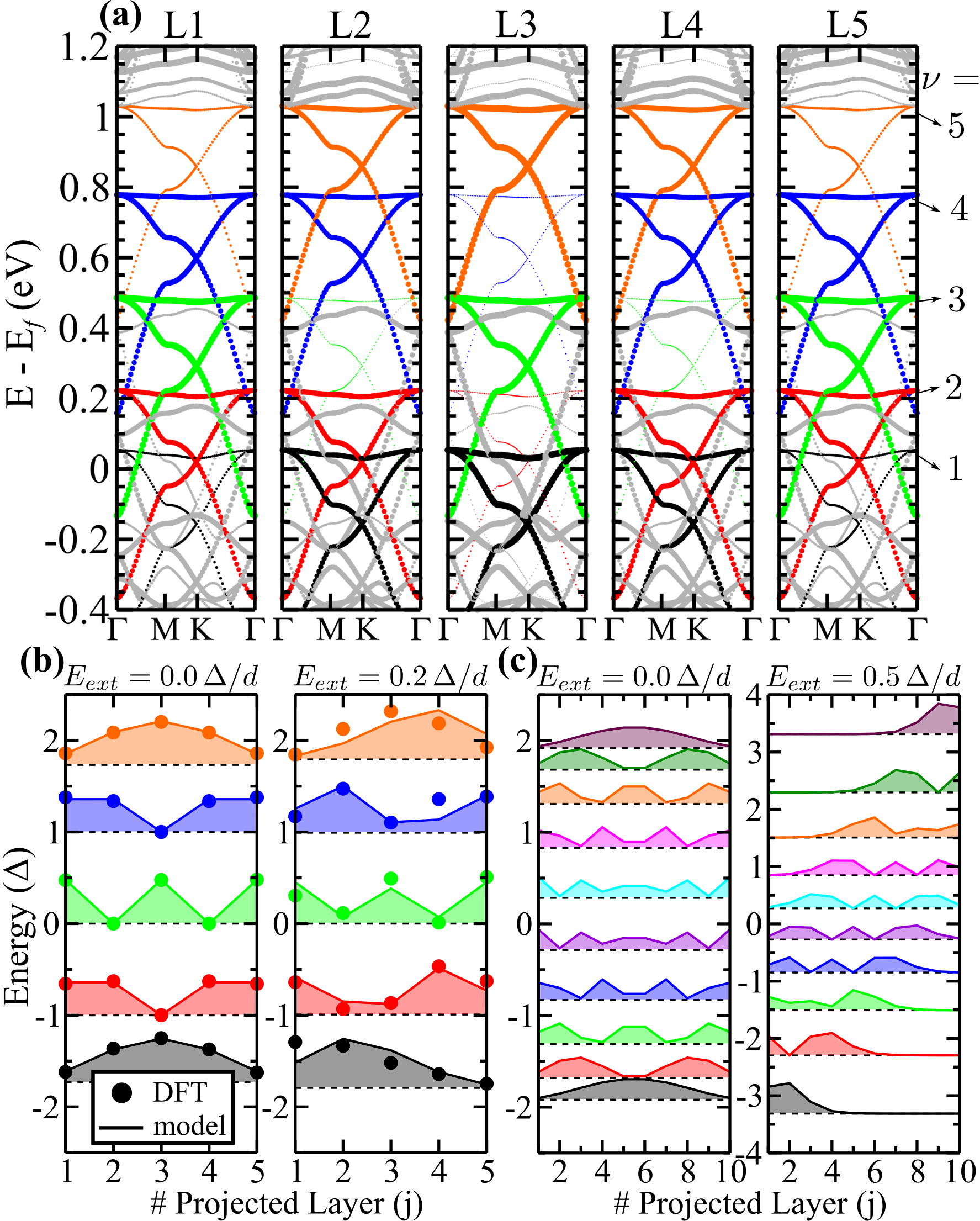}
\caption{\label{fig:localiz} (a) DFT band structure projected into each layer of a 5ML {\Ni} stacking. (b) Layer localization contributions $|\phi_{j,\nu}|^2$ extracted from the DFT and TB model without (with) external electric field in the left (right) panel. (c) Layer localization $|\phi_{j,\nu}|^2$ for a 10L slab from the TB model without (with) electric field in the left (right) panel.}
\end{figure}

The quantum well analogy, introduced above, enricher our discussion in the layer localization of the kagome states for finite AA stackings. First, in order to  verify the validity of this model, we performed {\it first-principles} DFT calculations of the KBs for a 5ML slab. Panels L1 to L5 in Fig.\,\ref{fig:localiz}(a) show the band structures projected on each layer $j$, and indicates the sets $\nu$ of KBs for clarity. Even though the band structure is dense at low energies, it is possible to see that: (i) the highest and the lowest KBs (${\nu\text{=1, and 5}}$) are mostly localized in the inner layers (L2, L3, L4); (ii) L3 does not contribute to ${\nu\text{=2, and 4}}$, which localizes mostly on the outer layers; (iii) the $\nu=3$ KBs contributes mostly to L1, L3 and L5. These features are more clearly seen in the circles of the left panel in Fig.\,\ref{fig:localiz}(b), which compares the layer projections $|\phi_{j,\nu}|^2$ extracted from the DFT data [Fig.\,\ref{fig:localiz}(a)], with the TB solutions. Indeed, these projections match remarkably well, thus validating the TB model.

In the presence of an external perpendicular electric field $\bm{E}_{ext} = E_{ext}\hat{z}$, the projections $|\phi_{j,\nu}|^2$ tend to shift to the outer layers. This is shown in Fig.\,\ref{fig:localiz}(b) and (c), for a 5ML and 10ML slab respectively. There, we compare the eigenstate $\phi_{j,\nu}$ distributions among the layers in the absence (left panels) and presence (right panels) of $E_{ext}$. For a 5ML slab, the DFT data on the right panel of Fig.~\ref{fig:localiz}(b) is obtained for a nominal $E_{ext} = 0.5$~eV/\AA. However, due to charge rearrangement between the layers, the actual electric field is inhomogeneous across the layers. Nonetheless, the DFT data matches qualitatively well with the TB model for an effective uniform field $E_{ext} = 0.2 \Delta/d \approx 0.02$~eV/\AA. Thus, the localization of each state follows the on-site energy shift due to $E_{ext}$, \textit{e.g.}, the lower energy KBs shifts towards L1. 

The same behavior is extrapolated to thicker slabs, for instance a 10ML slab as shown in Fig.\,\ref{fig:localiz}(c). Therefore $E_{ext}$ allows one to control the layer-localization of the KBs projection. In turn, combining this feature with the tunability of the Fermi energy (\textit{e.g.}, doping), one obtains an interesting engineering of the layer localization of topological edge states, which we discuss in the next section.

\subsection{Edge states}
\label{sec:edge}

The SOC opens an energy gap at the degenerated points ($\Gamma$ and K) of each set of KBs, yielding a QSH phase \cite{NANOFeng2013, PRBdeLima2017}. Interestingly, the weak vdW coupling between the layers of our {\Ni} AA stacking leads to a series of nearly independent edge states in each layer. To model this layered QSH phase, we add the SOC to $h_3(\bm{k})$ of our TB model, since the large number of atoms involved in ribbon geometries makes {\fp} calculations unattainable. The on-site, hopping, and SOC parameters within the single layer, and the $\Delta$ coupling between layers were obtained by fitting the {\it first-principles} DFT band structure of the bulk system. Further details can be found in the Appendix \ref{app:TB-par} and in Refs.\, \cite{PRBdeLima2017, PCCPdeLima2018}.

\subsubsection{Edge states on a finite stacking}

\begin{figure}[h!]
\includegraphics[width=0.95\columnwidth]{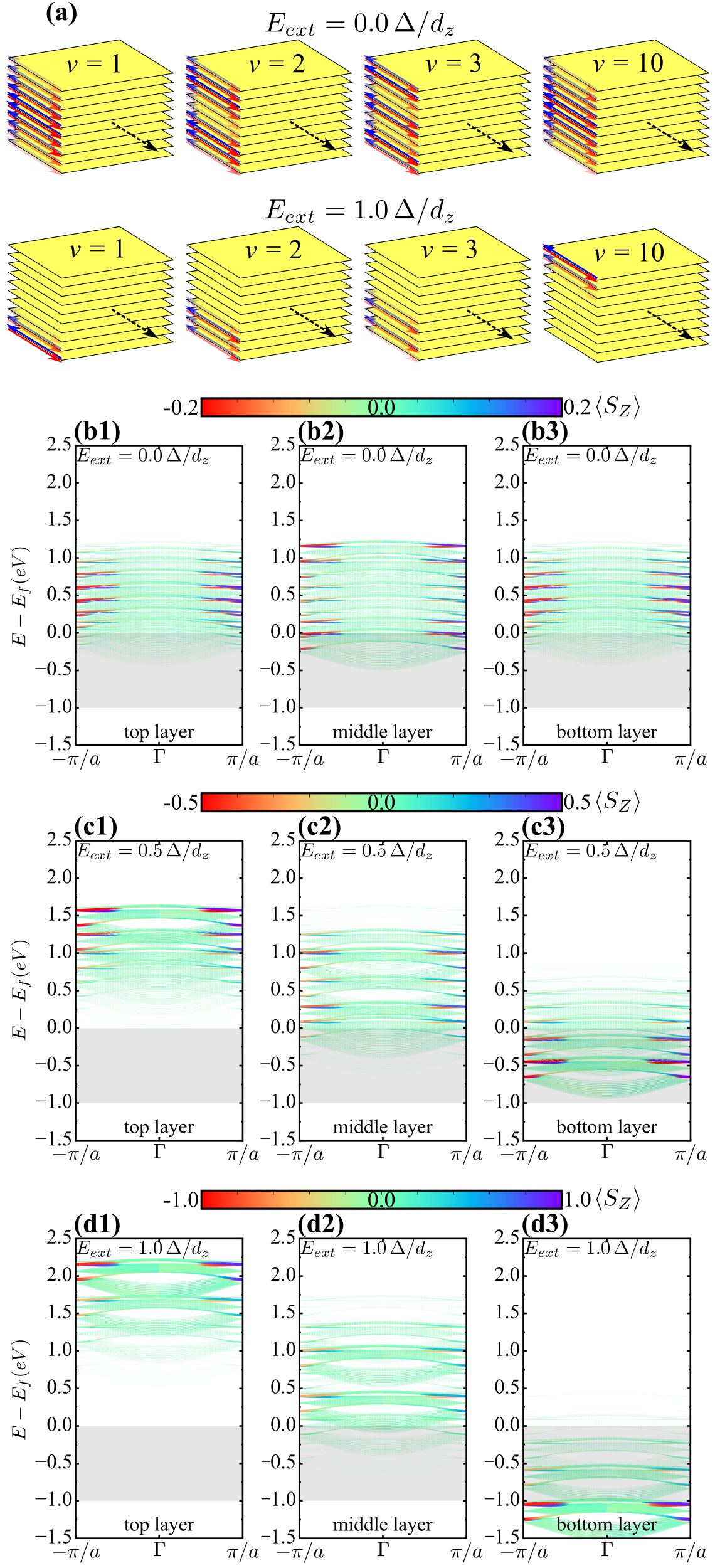}
\caption{\label{fig:tb-ribbon-ext} TB nanoribbon of multilayer {\Ni} with external electric field ($E_{ext}$). (a) visualization of edge states distribution for some  eigenvalues ($\nu$) without (with) $E_{ext}$ in the top (bottom) panels. Layered projected side surface states for $E_{ext} = 0.0$ (b), $=0.5$ (c) and $=1.0\,\Delta/d_z$ (d). Here, (b1)-(d1) is the projected band structure in the bottom most layer 1L, (b2)-(d2) in the middle 5L layer and (b3)-(d3) the top most layer 10L. The color bar indicate the $\langle S_z \rangle$ value and the size of the line is proportional to the layer contribution.}
\end{figure}

In this section we consider a 10ML AA stacking in a ribbon geometry. Nevertheless, the following discussion applies to any finite stacking. First, in the absence of electric fields, we calculate the spin-polarized edge states and present its projections into the topmost (L10), middle (L5), and bottommost (L1) layers in Figs.~\ref{fig:tb-ribbon-ext}(b1)--\ref{fig:tb-ribbon-ext}(b3). These projections agree with the layer resolved distribution of the KBs discussed previously. For instance, the $\nu=1$ state (low energy) is mostly located in the middle layer. The same is true for the $\nu=10$ (high energy) state, since their KBs projections $\phi_{j, \nu}$ are similar, see left panel of Fig.\,\ref{fig:localiz}(c). On the other hand, states within $2 \leq \nu \leq 9$ are spread along many layers, as illustrated in Fig.~\ref{fig:tb-ribbon-ext}(a). These results tell us that the vdW coupling between layers reflects into weakly coupled QSH edge states. Such edge states distribution allows for an unexploited control of the edge currents. For instance, tuning the Fermi level to the $\nu=1$ state, the helical currents will flow mostly through middle of the slab, while for $\nu=2$ state is distributed along its extremities, see top panel of Fig.\,\ref{fig:tb-ribbon-ext}(a).

On the other hand, in the presence of $E_{ext}$, the layer localization of the edge states follows that of the KBs, see right panel of Fig.\,\ref{fig:localiz}(c). This allows for a fine tuning of the edge states contributions near the Fermi level. To illustrate, Figs.~\ref{fig:tb-ribbon-ext}(c1)-(c3) show the edge state dispersions for an external field $E_{ext} = 0.5 \Delta/d_z \approx 0.05$\,eV/{\AA}. Here, we can see the lower (higher) energy edge states beginning to localize in the bottom (top) layer. Increasing the external field, $E_{ext} = 0.5 \rightarrow 1.0 \Delta/d_z$, the localization effect is enhanced, compare Figs,\,\ref{fig:tb-ribbon-ext}(c3) and (d3). That is, the field changes the on-site energies of each layer, pushing the $\nu=1$ state towards the bottom layer, as depicted in Fig.~\ref{fig:tb-ribbon-ext}(a). Consequently, the states near the Fermi level [shaded area in Figs.~\ref{fig:tb-ribbon-ext}(b)-(d).] are mostly located in the bottom layer. If the field is reversed, $E_{ext} \rightarrow -E_{ext}$, the states near the Fermi level will switch the localization to the top layer. This tuning of the edge states, by $E_{ext}$ and/or changing the Fermi level, allow for an additional degree of freedom that can be useful to design electronic (\textit{layertronic}) devices based on topologically protected edge currents in metal organic frameworks. 

\subsubsection{Edge states on the bulk stacking}

\begin{figure}
\includegraphics[width=\columnwidth]{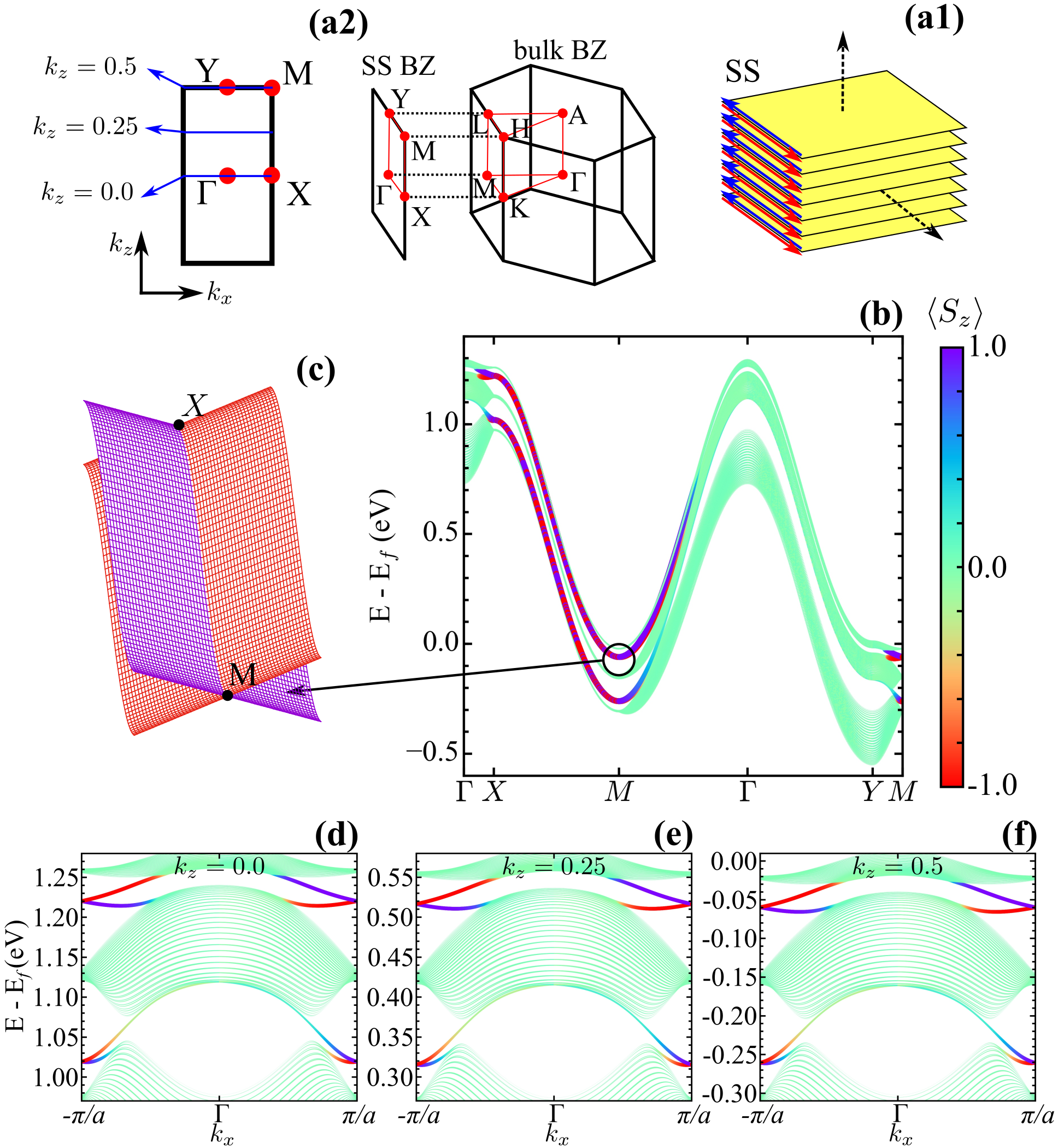}
\caption{\label{fig:tb-surface} Distribution of the side surface (SS) states (a1) where the dashed arrow indicate the periodic directions; bulk and SS Brillouin zone (a2); spin projected side surface states (b); zoom showing the band dispersion as a function of $(k_x,\,k_z)$ for the topological states close to the XM line (c). From (d) to (f) the spin projected side surface band structure for $k_z$ fixed varying $k_x$ as highlighted in blue in (a2). For the band plots the size of the lines are proportional to the sides surface sites contribution to the state $|\braket{{\rm SS}|n,\,{\bm k}}|^2$ while the color indicate the $S_z$ mean value.}
\end{figure}

The electronic confinement effects, present in the few layers systems, no longer takes place in the bulk phase. Here, the weakly coupled edge states from each layer are evenly spread along the side surfaces of the bulk staking, as shown schematically  in Fig.\,\ref{fig:tb-surface}(a1). Nonetheless, these side surface states carry the characteristics of the individual QSH edge states of the monolayers.

As discussed previously, the infinite superposition of the KBs gives rise to an continuum energy dispersion $\propto -2|\Delta|\cos(k_z d)$, where $k_z$ is the wave vector parallel to the stacking direction $\Gamma$-A. This bulk dispersion is projected in the 2D BZ of the side surface [Fig.\,\ref{fig:tb-surface}(a2) and \ref{fig:tb-surface}(b)], yielding a continuum of linear crossings at the edge of the Brillouin zone (X--M $\parallel k_z$), as shown in Fig.\,\ref{fig:tb-surface}(c).

Interestingly, this peculiar V-gutter-like edge state dispersion [Fig.\,\ref{fig:tb-surface}(c)] is highly anisotropic. An effective model for these states can be cast as a individual 2DTI edge states coupled along the stacking direction, wich reads
\begin{equation}
	E_{XM}(k_x, k_z) = \hbar v_F \sigma_z k_x - 2|\Delta| \sigma_0 \cos (k_z d),
\end{equation}
where $\sigma_0$ is the $2\times2$ identity matrix, $\sigma_z$ is third Pauli matrix in spin space, $v_F$ is the velocity of the edge states along $k_x$, and $\Delta$ is the same interlayer coupling from the bulk model. Along $k_x$, the states are helical with constant velocities $v_x = \pm v_F$ for spin up and down, while along $k_z$ the velocity is spin independent, but changes with $k_z$ or Fermi energy, $v_z = (|\Delta|d/\hbar)\sin(k_z d)$. This anisotropy is a consequence of the weak vdW coupling between the stacked layers. Thus, as shown in Fig.\,\ref{fig:bulk-strain}, the dynamics along $z$ can be further controlled by changing $\Delta$ with the uniaxial strain.

\section{Conclusion}

In this paper, we show that electronic layer localization can be engineered in multilayer MOFs by controlling the Fermi level (\textit{e.g.,} doping) and an external electric field $E_{ext}$. This is possible due to the van der Waals (vdW) coupling between adjacent layers on eclipsed AA stackings. Here, we consider the {\Ni} MOFs, showing that AA stacking is the most stable for this material. Nonetheless, our results remain valid for other MOFs that also allow for eclipsed stackings. The interlayer vdW coupling reflects into weakly coupled topological edge states in each monolayer of the stackings. Consequently, we have shown that the layer localization of edge states can be controlled by $E_{ext}$, applied along the stacking direction, thus providing a novel degree of freedom to the design of TI based electronics.

Furthermore, extrapolating the system to an infinite stacking of nanoribbons, the weak vdW coupling between layers leads to a new V-gutter-like Dirac states in the side surfaces. This new surface states are highly anisotropic, showing a trivial velocity along the stacking direction, while along the monolayer edge a helical topological current is always present for a wide range of energy.

\section{ACKNOWLEDGMENTS}

The authors acknowledge   financial   support   from   the Brazilian  agencies  
CNPq, and FAPEMIG, and the CENAPAD-SP and Laboratório Nacional de Computação Científica (LNCC-SCAFMat) for computer time.

\appendix

\section{Band structure for bilayer stacking}\label{app:band-stacking}
\begin{figure}[h!]
\includegraphics[width=\columnwidth]{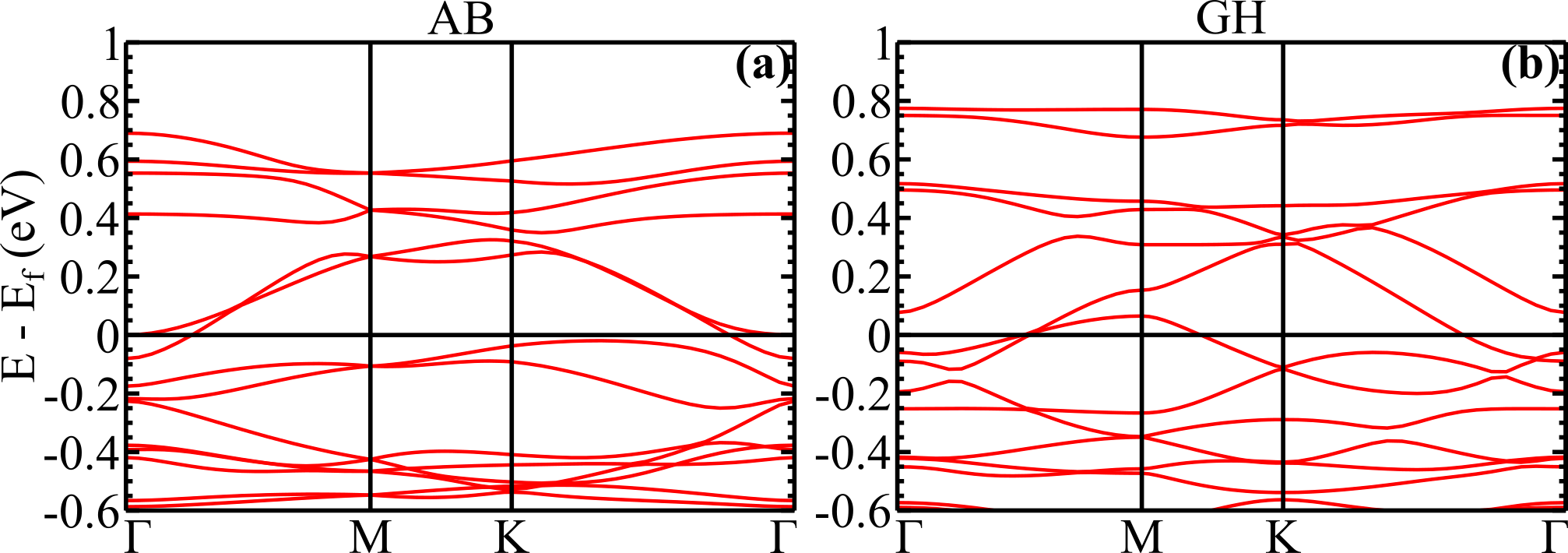}
\caption{\label{fig:sup-stk} Band structure for the {\Ni} bilayer in AB (a) and GH (b) stacking.}
\end{figure}

	For the {\Ni} bilayer, the stacking order has great impact in the electronic properties. For instance, in the mirror symmetric AA stacking, Fig.\,\ref{fig:band}(a2), we find sets of the characteristc kagome band dispersion. On the other hand, for the non-mirror symmetric stacking, Fig.\,\ref{fig:sup-stk}(a) and (b), the coupling between the layers leads to a different hybridization between the states, where the characteristic kagome bands are no longer present.
	
\begin{figure}
\includegraphics[width=\columnwidth]{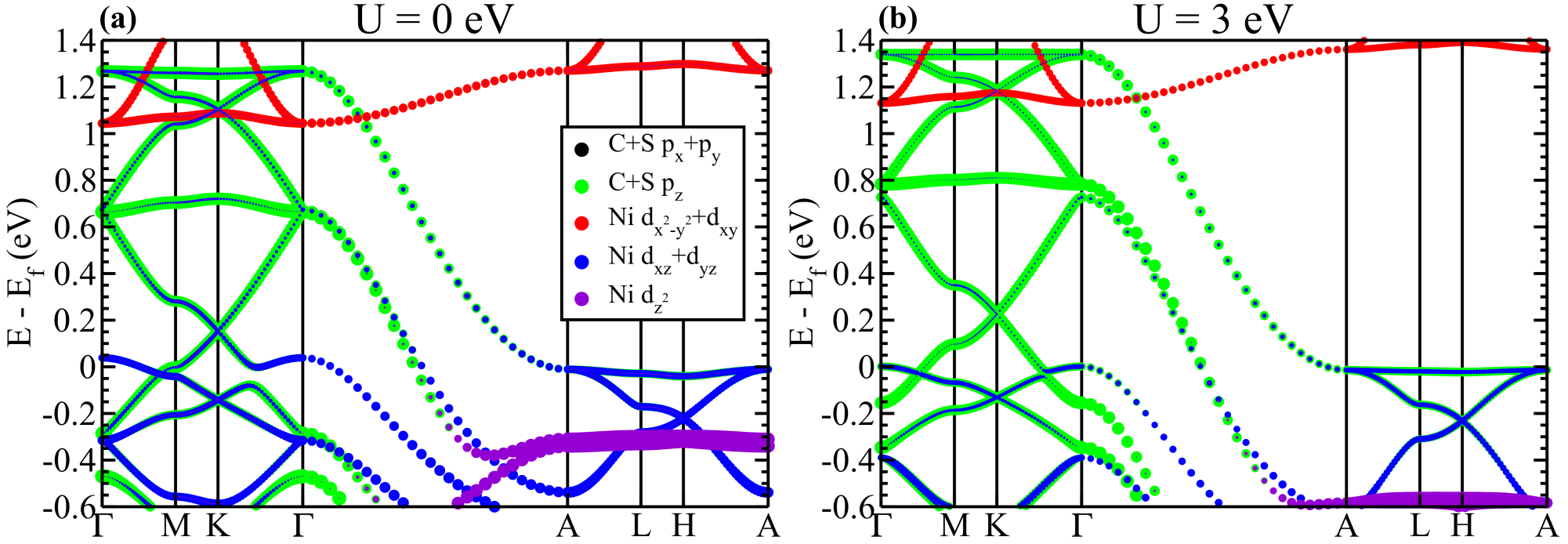}
\caption{\label{fig:sup:dft+u} { DFT+U calculation for the (NiC$_4$S$_4$)$_3$ bulk, considering the Hubbard on-site repulsion in the Ni d orbitals with $U=0$\,eV (a) and $U=3$\,eV (b).}}
\end{figure}

{\section{DFT+U}\label{app:DFT+U}
	We have performed GGA+U calculations {\cite{DFTU}} in order to verify the effects of the Hubbard on-site repulsion term for the d-orbitals of Ni in {\Ni}. We find that our main results are not affected by this correction. The comparison between the $U=0$\,eV and $U=3$\,eV cases in Fig.\,\ref{fig:sup:dft+u}(a) and (b) shows that the Ni d orbitals goes slightly down in energy for a finite $U$, which gives a negligible change in the charge state of Ni atoms, going from $+0.66\,e$ for $U=0$ to $+0.61\,e$ for $U=3$\,eV. Nevertheless, Fig.\,\ref{fig:sup:dft+u} shows that the Kagome band dispersions are not significantly affected by $U$. Since this is the main characteristic that leads to the tunability of the layer contributions (the sets of kagome bands), we can conclude that our results are robust, and the Hubbard repulsion term can be neglected here.
}	

\section{TB model parameters from DFT}\label{app:TB-par}

\begin{figure}
\includegraphics[width=\columnwidth]{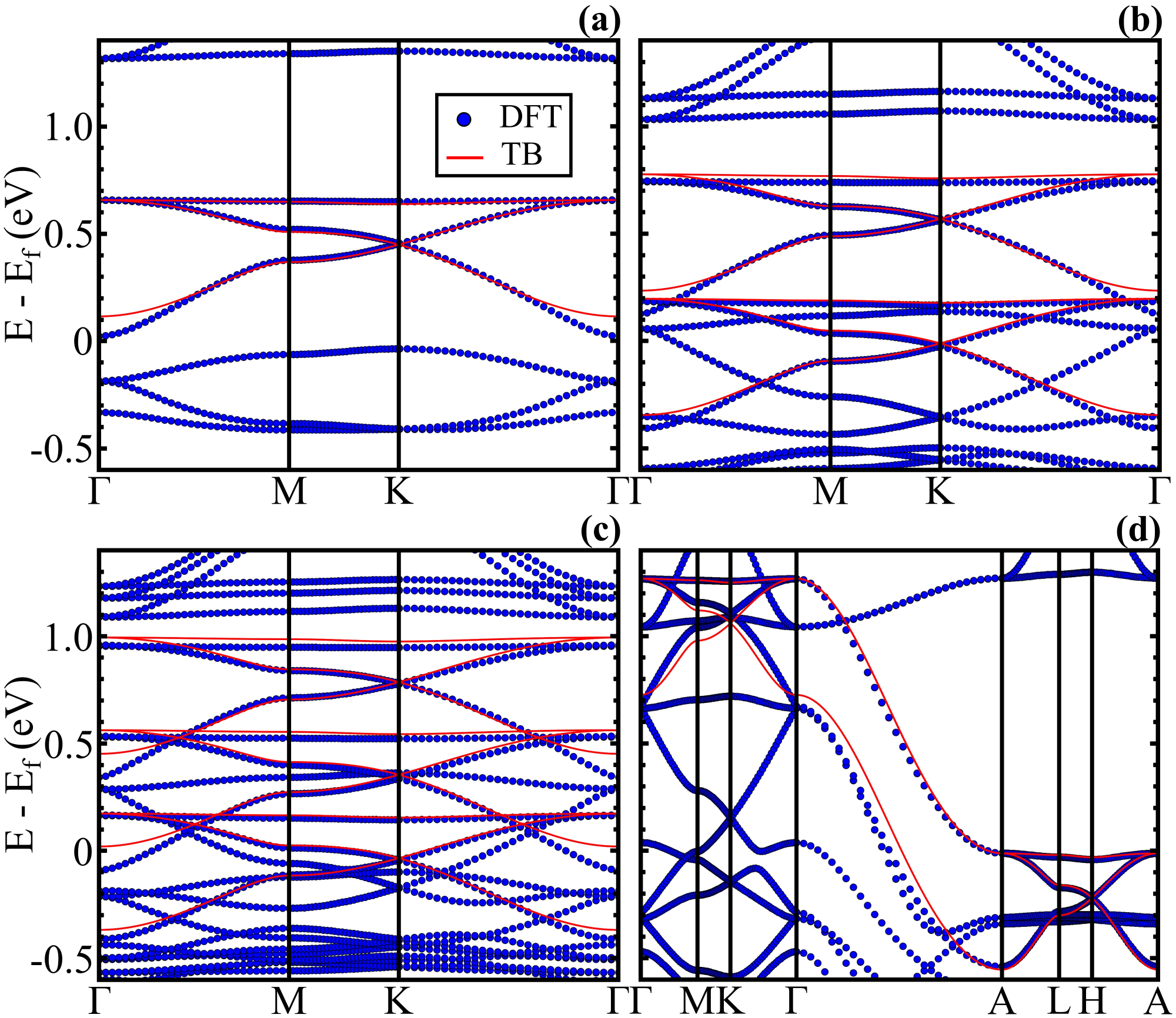}
\caption{\label{fig:sup:tb-dft-fit} TB fitting of the DFT band structures for 1L (a), 2L (b), 3L (c), and bulk (d).}
\end{figure}

	The considered tight-binding (TB) Hamiltonian is given by
\begin{equation}
H = H_0 + H_{SO} + H_{L},
\end{equation}
where $H_0$ contains the on-site energy $\epsilon_0$ and hoppings within each layer $L$,
\begin{equation}
H_0 = \sum_{L} \left[ \sum_{i\in L} \epsilon_0 \, c_i^{\dagger}c_i + \sum_{(i\neq j)\in L} t_{ij}\, c_i^{\dagger} c_j \right],
\end{equation}
with $i$ and $j$ indicating sites in the same layer, and sum over $L$ runs over all layers. $H_{SO}$ is the intrinsic spin-orbit coupling,
\begin{equation}
H_{SO} = \sum_{L} \left[ i \sum_{(i\neq j)\in L} \lambda_{ij}\, c_i^{\dagger} {\bm \sigma} \cdot \frac{\left( {\bm d}_{kj} \times {\bm d}_{ik} \right)}{|d_{kj}||d_{ik}|} c_j \right].
\end{equation}
The last term, $H_{L}$ contains the coupling between layers,
\begin{equation}
H_{L} = \sum_{L\neq L'} \left[ \sum_{k\in L} \sum_{l \in L'} t^z_{kl} \, c_{k}^{\dagger} c_l \right],
\end{equation}
where $k$ and $l$ indicates the same site, but in different layers. Here, $c_i^{\dagger} = (c_{i\uparrow}^{\dagger},c_{i\downarrow}^{\dagger})^T$, $c_i = (c_{i\uparrow},ci_{i\downarrow})^T $, $c_{i\chi}^{\dagger}$ ($c_{i\chi}$)  is the creation (annihilation) operators of an electron in the $i$-th site with spin $\chi$, and ${\bm \sigma}$ are the spin Pauli matrices. The ${\bm d}_{ij}$  vectors connect the $i$-th to the $j$-th site. For the hoppings ($t_{ij}$ and $t_{ij}^z$) and spin-orbit couplings ($\lambda_{ij}$), we consider a  distance dependent parametrization, which allow us to control the hopping range. This parametrization reads
\begin{equation}
t_{ij} = N\,t\,\exp ( -\alpha d_{ij} ),
\end{equation}
\begin{equation}
t_{ij}^z = N_z\,\Delta\,\exp (-\alpha_z d_{ij} ),
\end{equation}
and
\begin{equation}
\lambda_{ij} = N\,\lambda\,\exp (-\alpha d_{ij} ).
\end{equation}
All hoppings are normalized in relation to the first neighbor hopping, by the factor $N=\exp ( \alpha \,d_{nn} )$ and $N_z = \exp ( \alpha_z\, d_z )$, where $d_z$ is the interlayer distance. In Fig.\,\ref{fig:sup:tb-dft-fit} we show a comparison between the DFT and TB model, using the parameters shown in Tab.\,\ref{tab:sup:tb-par}. {Note that the TB parameter $\Delta$ is consistent with the one obtained through the analytic model.}

\begin{table}
\caption{\label{tab:sup:tb-par} TB parameters used to fit the DFT band structure of Fig.\,\ref{fig:sup:tb-dft-fit}.}
\begin{ruledtabular}
\begin{tabular}{cccccc}
 System  &   $\epsilon_0$   &   $t$   &   $\alpha$=$\alpha_z$  & $\lambda$  & $\Delta$  \\
\hline
  1L     &      0.50        &  -0.08  &     3.0     & 0.0 &   -    \\
  2L     &      0.33        &  -0.08  &     3.0     & 0.0 & -0.29 \\
  3L     &      0.42        &  -0.08  &     3.0     & 0.0 & -0.29 \\
 bulk    &      0.44        &  -0.08  &     3.0     & 0.0 & -0.32 \\
\end{tabular}
\end{ruledtabular}
\end{table}

\bibliography{bib}

\end{document}